\documentclass[a4paper,twoside]{article}

\usepackage[T1]{fontenc}
\usepackage[latin9]{inputenc}
\usepackage{graphicx}
\usepackage{amssymb}

\def\araa{ARA\&A}             % Annual Review of Astron and Astrophys
\def\apj{ApJ}                 % Astrophysical Journal
\def\apjl{ApJ}                % Astrophysical Journal, Letters
\def\apjs{ApJS}               % Astrophysical Journal, Supplement
\def\aap{A\&A}                % Astronomy and Astrophysics
\def\mnras{MNRAS}             % Monthly Notices of the RAS
\def\pasa{PASA}               % Publications of the ASP
\def\pasj{PASJ}               % Publications of the ASJ
\newcommand{\affil}[1]{$^{\rm #1}$}
\usepackage[authoryear]{natbib}
\bibpunct{(}{)}{;}{a}{}{,}

\date{} %Please leave the date blank
\title{\large\bf\flushleft Binary Populations of Carbon-Enhanced Metal-Poor Stars}
\author{\parbox{\textwidth}{\flushleft
\vspace{-0.5cm}
{\it Robert G. Izzard\affil{A,B}, Evert Glebbeek\affil{B,C}, Richard J. Stancliffe\affil{D,E}, and \\Onno Pols\affil{B}}\\
\vspace{0.4cm}
{\small \affil{A}\,Institut d'Astronomie et d'Astrophysique, Universit{\'e}
Libre de Bruxelles, Boulevard du Triomphe, B-1050 Brussels, Belgium.
Email: Robert.Izzard@ulb.ac.be}\\
{\small \affil{B}\,Sterrenkundig Instituut, Universiteit Utrecht, P.O.
Box 80000, NL-3508 TA Utrecht, The Netherlands.}\\
{\small \affil{C}\,Department of Physics and Astronomy, McMaster University,
Hamilton, Ontario, L8S 4M1, Canada.}\\
{\small \affil{D}\,School of Mathematical Sciences, PO Box 28M, Monash
University, Victoria 3800, Australia.}\\
{\small \affil{E}\,Institute of Astronomy, University of Cambridge,
Madingley Road, Cambridge, CB3 0HA, United Kingdom.}}}

\begin{document}
\twocolumn[
\begin{changemargin}{.8cm}{.5cm}
\begin{minipage}{.9\textwidth}
\vspace{-1cm}
\maketitle
%
%
%%%%%%%%%%%%%     ABSTRACT    %%%%%%%%%%%%%
%Abstract of no more than 200 words here.
\small{\bf Abstract:} We construct binary-star population nucleosynthesis models of carbon-enhanced
metal poor (CEMP) stars. We compare the CEMP to EMP (extremely metal
poor) ratio of our models to the observed ratio and find it is an
order of magnitude too small. Through an increase in the efficiency
of third dredge-up in low-mass, low-metallicity thermally-pulsing
asymptotic-giant branch (TPAGB) stars our models better match the
observations.

%%%%%%%%%%%%%     KEYWORDS    %%%%%%%%%%%%%
\medskip{\bf Keywords:} Binary stars --- mass transfer --- Galactic halo --- metal-poor stars
% Please write all keywords in lower case. PASA uses the
% standard list of subject headings adopted by The Astrophysical Journal
% and available from http://www.journals.uchicago.edu/ApJ/keywords_text.html.
% Keywords are separated by em-dashes, i.e. ---

%%%%%%%%DO NOT EDIT%%%%%%%%%%%%
\medskip
\medskip
\end{minipage}
\end{changemargin}
]
\small
%%%%%%%%EDIT FROM HERE%%%%%%%%%%%%

\section{Introduction}
%Please see the PASA Style Guide for help with correct layout for your manuscript.
%Examples of tables and figures are given below.

The carbon-enhanced metal-poor stars (CEMPs) are Galactic halo objects
which are observed to be iron poor ($\mathrm{\left[\mathrm{Fe}/\mathrm{H}\right]}\lesssim-2$),
relatively carbon rich ($[\mathrm{C}/\mathrm{Fe}]\geq1$) and mostly
giant and turn-off stars ($\log g\lesssim4$). Their mass is around
$0.85\mathrm{\, M_{\odot}}$, as dictated by the age of the halo.
Canonical models of $\sim0.85\mathrm{\, M_{\odot}}$ single stars
do not predict such a large carbon enhancement at any phase of their
evolution. Instead, it seems likely that most CEMPs form in binary
systems as a result of mass transfer from a carbon-rich thermally-pulsing
asymptotic giant branch (TPAGB) star which is now an unseen white
dwarf. Radial velocity surveys concur, at least for those CEMPs which
are also rich in $s$-process elements (\citealp*{2004MmSAI..75..772T};
\citealp{2005ApJ...625..825L}). According to detailed TPAGB models
the primary star should have had a mass in the range $1.2$ to $3\mathrm{\, M_{\odot}}$,
in order to undergo efficient third dredge-up of carbon without hot-bottom
burning (which converts carbon to nitrogen, \citealp*[e.g.][]{2007PASA...24..103K}). For a review of the properties of carbon stars see e.g. \citet{1998ARA&A..36..369W} or \citet{2003PASA...20..314A}.

The CH stars are thought to be the higher metallicity ($\mathrm{\left[\mathrm{Fe}/\mathrm{H}\right]}\sim-1$)
equivalents of the CEMPs because they form by a similar binary-accretion
mechanism. They make up about $1\%$ of the giant population \citep{1991ApJS...77..515L},
in rough agreement with our theoretical estimates based on binary
population synthesis (see Section~\ref{sec:models}). By contrast
9--30\% of the EMP (extremely metal poor) giant population are CEMPs (\citealp{2006ApJ...652.1585F}; \citealp{2006ApJ...652L..37L};
\citealp{2008arXiv0806.3697S}), in disagreement with our standard
binary population models. 

Recent studies suggest that an initial-mass function (IMF) quite different
to that of the solar neighbourhood is responsible for the large CEMP
to EMP (extremely metal poor) number ratio (e.g. \citealp{2005ApJ...625..833L}; 
\citealp{2007ApJ...658..367K}). We attempt to explain the CEMP to
EMP number ratio without a change in the initial distributions of
stellar masses and initial periods from those found in the solar neighbourhood. 

Many of the physical parameters in our binary-star model are quite
uncertain, but we show that most are not important in calculation
of the CEMP to EMP number ratio. The parameters which matter most
are those that affect the efficiency of third dredge-up in low-mass
stars.

\section{Binary Population Models}

\label{sec:models}We base our population synthesis models on the
rapid binary-star evolution and nucleosynthesis models of \citet*{2002MNRAS_329_897H};
\citet{Izzard_et_al_2003b_AGBs} and \citet{2006A&A...460..565I}.
Our nucleosynthesis algorithm has been updated~~to~~better~~model first
dredge-up in low-metallicity stars which have accreted carbon-rich
material from a companion. Third dredge-up is modelled as in \citet*{Parameterising_3DUP_Karakas_Lattanzio_Pols}
using the parameters $M_{\mathrm{c},\mathrm{min}}$, the minimum core
mass for third dredge-up, and $\lambda$, , the ratio of the mass of
material which is dredged up at a given pulse to the mass by which
the core grew during the preceding interpulse period.

We also include the correction factors $\Delta M_{\mathrm{c},\mathrm{min}}$
and $\lambda_{\mathrm{min}}$ which were introduced in \citet{Izzard_et_al_2003b_AGBs}
in order to match our synthetic TPAGB models to observed carbon-star
luminosity functions in the Magellanic Clouds. These alter the prescription
of \citet*{Parameterising_3DUP_Karakas_Lattanzio_Pols}~~such~~that
$M_{\mathrm{c},\mathrm{min}}=M_{\mathrm{c},\mathrm{min}}^{\mathrm{Karakas}}+\Delta M_{\mathrm{c},\mathrm{min}}$
and $\lambda=\max\left(\lambda^{\mathrm{Karakas}},\lambda_{\mathrm{min}}\right)$.

A new parameter in our model is $M_{\mathrm{env},\mathrm{min}}$,
the minimum envelope mass a star must have in order for third dredge-up
to occur. This was set to $0.5\mathrm{\, M_{\odot}}$ in previous
models, following \citet{1997ApJ...478..332S}, but will be treated
as a free parameter for our purposes. Note that while our models include
canonical third dredge-up and hot-bottom burning, they do not contain
prescriptions for dual-core flashes or dual-shell flashes (e.g. \citealp{2007arXiv0709.4567C},
Cristallo et al. this volume) which are expected to occur at low metallicity
($[\mathrm{Fe}/\mathrm{H}]\lesssim-3$). 
\vspace{2mm}
Our standard choice of physical parameters is described in some detail
in \citet{2006A&A...460..565I} and the metallicity is, in most of
our models, $Z=10^{-4}$ (such that $[\mathrm{Fe}/\mathrm{H}]=-2.3$). We distribute
initial primary masses according to the IMF of \citet*{KTG1993MNRAS-262-545K},
the binary mass ratio is distributed evenly between $0$ and $1$
and the separation distribution is flat in $\log$--separation between
$3$ and $10^{5}\mathrm{\, R_{\odot}}$ (the upper limit is chosen
to include all CEMPs formed by wind accretion). We assume a binary
fraction of $100\%$.

\begin{figure*}
\begin{centering}
\includegraphics[angle=270,scale=0.45]{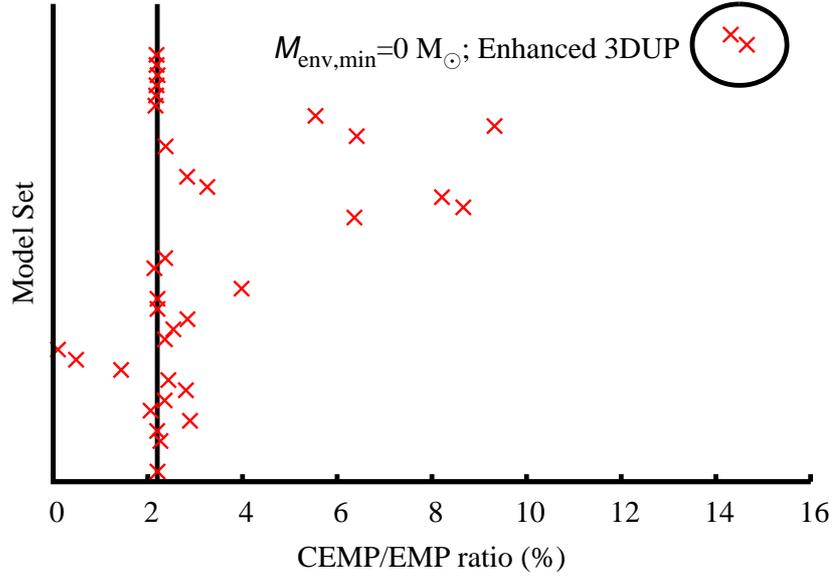}
\par

\caption{\label{fig:The-CEMP/EMP-ratio}The CEMP/EMP ratio for our simulated
binary populations. The abscissa is the CEMP/EMP ratio while the ordinate
is chosen arbitrarily to separate the models with different physical
parameters. The black vertical line at $2.2\%$ shows the CEMP/EMP
ratio of our default model set.\vspace{-4mm}}
\end{centering}
\end{figure*}

\vspace{2mm}
We constructed dozens of binary-star populations with a wide variety
of physical parameter choices, as shown in Figure~\ref{fig:The-CEMP/EMP-ratio}.
These include alternative TPAGB wind prescriptions, enhanced wind
accretion efficiency, tidally-enhanced~~mass~~loss,~~alternative common-envelope
evolution prescriptions, accretion of material in a common envelope
phase, lower initial metallicity and various combinations of $\Delta M_{\mathrm{c},\mathrm{min}}$,
$\lambda_{\mathrm{min}}$ and $M_{\mathrm{env},\mathrm{min}}$. Figure~\ref{fig:The-CEMP/EMP-ratio}
shows that most combinations of parameters do not lead to a CEMP/EMP
ratio which is anywhere near the observed values -- instead most results
cluster around a ratio of 2\% (similar to the CH-to-giant ratio mentioned
above). We conclude that most\emph{ }of the free parameters in our
model are not important for CEMP formation.

\vspace{2mm}
This conclusion is untrue for the parameters which affect third dredge-up
in low-mass stars: $\Delta M_{\mathrm{c},\mathrm{min}}$, $\lambda_{\mathrm{min}}$
and $M_{\mathrm{env},\mathrm{min}}$. A choice of $\Delta M_{\mathrm{c},\mathrm{min}}\sim-0.1\mathrm{\, M_{\odot}}$
reduces the minimum core mass for third dredge-up so that it can occur
in low-mass stars right down to the lower-mass limit of $0.85\mathrm{\, M_{\odot}}$.
A positive $\lambda_{\mathrm{min}}$ forces third dredge-up to be
efficient in low-mass stars (we chose values of $0.5$ and greater).
Also important is $M_{\mathrm{env},\mathrm{min}}$: the default value
of $0.5\mathrm{\, M_{\odot}}$ prevents third dredge-up in low-mass
stars because they have a small envelope mass that is too small. When
we set $M_{\mathrm{env},\mathrm{min}}$ to zero, together with $\Delta M_{\mathrm{c},\mathrm{min}}\sim-0.1\mathrm{\, M_{\odot}}$
and $\lambda_{\mathrm{min}}\sim0.5$, we find CEMP/EMP ratios of up
to 15\%. These ratios are far more compatible with the observations.

\vspace{2mm}
We must point out that we have artificially increased the amount of
third dredge-up in low-mass TPAGB stars without any physical justification.
However, new detailed stellar evolution models by \citet{2008arXiv0807.1758S}
show third dredge-up in a $0.9\mathrm{\, M_{\odot}}$ star, albeit
with an efficiency of only $\lambda=0.16$. Alternatively, Cristallo
et al. (this volume) show that proton ingestion at a first {}``huge
pulse'' may occur at higher metallicities than previously thought,
perhaps up to $[\mathrm{Fe}/\mathrm{H}]=-2$, in stars of mass around
$0.85\mathrm{\, M_{\odot}}$. This mechanism also leads to an efficient
third dredge-up and carbon enrichment of the stellar envelope, as
our models suggest is required. Work to simulate populations of binaries
which include this mechanism is underway.\vspace{-1.5mm}

\section{Tentative Conclusions}

If TPAGB stars with masses as low as $0.85\mathrm{\, M_{\odot}}$
and metallicity around $[\mathrm{Fe/H]\approx-2.3}$ undergo efficient
third dredge-up they may be responsible for the formation of the majority
of the CEMP stars. Canonical detailed TPAGB models do not tend to
support this conclusion, as only stars with masses above about $1.2\mathrm{\, M_{\odot}}$
show third dredge-up.

New detailed models suggest that third dredge-up may occur in low-mass,
low-metallicity stars, possibly due to a {}``huge first pulse''.
Even if we include this very efficient third dredge-up in low-mass
stars in our binary population models, we cannot increase the CEMP
to EMP number ratio beyond $15\%$ even with a binary fraction of
$100\%$. Perhaps a combination of efficient third dredge-up with
alternative binary distributions (the IMF, mass-ratio and period distributions)
is responsible (Pols et al., this volume).

\section*{Acknowledgments} %If needed
RGI thanks the NWO for his fellowship in Utrecht and is the recipient of a Marie Curie-Intra European Fellowship at ULB. RJS is funded by the Australian Research Council's Discovery Projects scheme under grant DP0879472. He is grateful to Churchill College for his Junior Research Fellowship, under which this work commenced.


\begin{thebibliography}{}

\bibitem[{Campbell} and {Lattanzio}, 2008]{2007arXiv0709.4567C}
{Campbell}, S.~W. and {Lattanzio}, J.~C. (2008).
\newblock {Structural and Nucleosynthetic Evolution of Metal-poor and
  Metal-free Low and Intermediate Mass Stars}.
\newblock American Institute of Physics Conference Series 990:315--319.

\bibitem[{Frebel} et~al., 2006]{2006ApJ...652.1585F}
{Frebel}, A., {Christlieb}, N., {Norris}, J.~E., {Beers}, T.~C., {Bessell},
  M.~S., {Rhee}, J., {Fechner}, C., {Marsteller}, B., {Rossi}, S., {Thom}, C.,
  {Wisotzki}, L., and {Reimers}, D. (2006).
\newblock {Bright Metal-poor Stars from the Hamburg/ESO Survey. I. Selection
  and Follow-up Observations from 329 Fields}.
\newblock {\em \apj}, 652:1585--1603.

\bibitem[{Hurley} et~al., 2002]{2002MNRAS_329_897H}
{Hurley}, J.~R., {Tout}, C.~A., and {Pols}, O.~R. (2002).
\newblock {Evolution of binary stars and the effect of tides on binary
  populations}.
\newblock {\em \mnras}, 329:897--928.

\bibitem[{Izzard} et~al., 2006]{2006A&A...460..565I}
{Izzard}, R.~G., {Dray}, L.~M., {Karakas}, A.~I., {Lugaro}, M., and {Tout},
  C.~A. (2006).
\newblock {Population nucleosynthesis in single and binary stars. I. Model}.
\newblock {\em \aap}, 460:565--572.

\bibitem[{Izzard} et~al., 2004]{Izzard_et_al_2003b_AGBs}
{Izzard}, R.~G., {Tout}, C.~A., {Karakas}, A.~I., and {Pols}, O.~R. (2004).
\newblock {A New Synthetic Model for AGB Stars}.
\newblock {\em \mnras}, 350:407--426.

\bibitem[{Karakas} and {Lattanzio}, 2007]{2007PASA...24..103K}
{Karakas}, A. and {Lattanzio}, J.~C. (2007).
\newblock {Stellar Models and Yields of Asymptotic Giant Branch Stars}.
\newblock {\em PASA},
  24:103--117.

\bibitem[{Karakas} et~al., 2002]{Parameterising_3DUP_Karakas_Lattanzio_Pols}
{Karakas}, A.~I., {Lattanzio}, J.~C., and {Pols}, O.~R. (2002).
\newblock {Parameterising the third dredge-up in asymptotic giant branch
  stars.}
\newblock {\em \pasa}, 19:515--526.

\bibitem[{Komiya} et~al., 2007]{2007ApJ...658..367K}
{Komiya}, Y., {Suda}, T., {Minaguchi}, H., {Shigeyama}, T., {Aoki}, W., and
  {Fujimoto}, M.~Y. (2007).
\newblock {The Origin of Carbon Enhancement and the Initial Mass Function of
  Extremely Metal-poor Stars in the Galactic Halo}.
\newblock {\em \apj}, 658:367--390.

\bibitem[{Kroupa} et~al., 1993]{KTG1993MNRAS-262-545K}
{Kroupa}, P., {Tout}, C., and {Gilmore}, G. (1993).
\newblock {The distribution of low-mass stars in the Galactic disc}.
\newblock {\em \mnras}, 262:545--587.

\bibitem[{Lucatello} et~al., 2006]{2006ApJ...652L..37L}
{Lucatello}, S., {Beers}, T.~C., {Christlieb}, N., {Barklem}, P.~S., {Rossi},
  S., {Marsteller}, B., {Sivarani}, T., and {Lee}, Y.~S. (2006).
\newblock {The Frequency of Carbon-enhanced Metal-poor Stars in the Galaxy from
  the HERES Sample}.
\newblock {\em \apjl}, 652:L37--L40.

\bibitem[{Lucatello} et~al., 2005a]{2005ApJ...625..833L}
{Lucatello}, S., {Gratton}, R.~G., {Beers}, T.~C., and {Carretta}, E. (2005a).
\newblock {Observational Evidence for a Different Initial Mass Function in the
  Early Galaxy}.
\newblock {\em \apj}, 625:833--837.

\bibitem[{Lucatello} et~al., 2005b]{2005ApJ...625..825L}
{Lucatello}, S., {Tsangarides}, S., {Beers}, T.~C., {Carretta}, E., {Gratton},
  R.~G., and {Ryan}, S.~G. (2005b).
\newblock {The Binary Frequency Among Carbon-enhanced, s-Process-rich,
  Metal-poor Stars}.
\newblock {\em \apj}, 625:825--832.

\bibitem[{Luck} and {Bond}, 1991]{1991ApJS...77..515L}
{Luck}, R.~E. and {Bond}, H.~E. (1991).
\newblock {Subgiant CH stars. II - Chemical compositions and the evolutionary
  connection with barium stars}.
\newblock {\em \apjs}, 77:515--540.

\bibitem[{Stancliffe} and {Glebbeek}, 2008]{2008arXiv0807.1758S}
{Stancliffe}, R.~J. and {Glebbeek}, E. (2008).
\newblock {Thermohaline mixing and gravitational settling in carbon-enhanced
  metal-poor stars}.
\newblock {\em \mnras}, 389:1828--1838.

\bibitem[{Straniero} et~al., 1997]{1997ApJ...478..332S}
{Straniero}, O., {Chieffi}, A., {Limongi}, M., {Busso}, M., {Gallino}, R., and
  {Arlandini}, C. (1997).
\newblock {Evolution and Nucleosynthesis in Low-Mass Asymptotic Giant Branch
  Stars. I. Formation of Population I Carbon Stars}.
\newblock {\em \apj}, 478:332.

\bibitem[{Suda} et~al., 2008]{2008arXiv0806.3697S}
{Suda}, T., {Katsuta}, Y., {Yamada}, S., {Suwa}, T., {Ishizuka}, C., {Komiya},
  Y., {Sorai}, K., {Aikawa}, M., and {Fujimoto}, M.~Y. (2008).
\newblock {The Stellar Abundances for Galactic Archeology (SAGA) Database -
  Compilation of the Characteristics of Known Extremely Metal-Poor Stars}.
\newblock {\em \pasj}, 60:1159.

\bibitem[{Tsangarides} et~al., 2004]{2004MmSAI..75..772T}
{Tsangarides}, S., {Ryan}, S.~G., and {Beers}, T.~C. (2004).
\newblock {On the binarity of carbon-enhanced, metal-poor stars}.
\newblock {\em Memorie della Societa Astronomica Italiana}, 75:772.


\bibitem[Wallerstein and Knapp, 1998]{1998ARA&A..36..369W}
{Wallerstein}, G. and {Knapp}, G.~R. (1998)
\newblock {Carbon Stars}.
\newblock {\em \araa}, 36:369.

\bibitem[{Abia} et~al., 2003]{2003PASA...20..314A}
{Abia}, C., {Dom{\'{\i}}nguez}, I., {Gallino}, R., {Busso}, M., {Straniero}, O., {de Laverny}, P. and {Wallerstein}, G.
\newblock {Understanding AGB Carbon Star Nucleosynthesis from Observations}.
\newblock {\em PASA}, 20:314


\end{thebibliography}
\end{document}